\documentstyle[11pt]{article} 
\def\cF{\cal F}

\newfont{\goth}{eufm10 scaled \magstep1}

\def\a{\alpha}

\def\b{\beta}
\def\db{{\dot\beta}}
\def\c{\gamma}\def\C{\Gamma}
\def\d{\delta}
\def\e{\epsilon}

\def\h{\eta}
\def\k{\kappa}
\def\l{\lambda}
\def\m{\mu}

\def\s{\sigma}\def\S{\Sigma}
\def\t{\tau}
\def\th{\theta}

\def\beq{\begin{equation}}\def\eeq{\end{equation}}
\def\beqa{\begin{eqnarray}}\def\eeqa{\end{eqnarray}}
\def\barr{\begin{array}}\def\earr{\end{array}}
\def\x{\xi}
\def\o{\omega}
\def\del{\partial}
\def\ua{\underline{\alpha}}
\def\ub{\underline{\phantom{\alpha}}\!\!\!\beta}
\def\uc{\underline{\phantom{\alpha}}\!\!\!\gamma}

\def\una{\underline a}\def\unA{\underline A}
\def\unb{\underline b}\def\unB{\underline B}
\def\unc{\underline c}\def\unC{\underline C}
\def\und{\underline d}

\def\unm{\underline m}\def\unM{\underline M}

\let\la=\label

\let\bm=\bibitem

\def\uM{{\underline M}}
\def\umu{{\underline \mu}}
\def\nn{\nonumber}
\def\bd{\begin{document}}
\def\ed{\end{document}}
\def\ba{\begin{array}}
\def\ea{\end{array}}
\def\bea{\begin{eqnarray}}
\def\eea{\end{eqnarray}}
\def\ft#1#2{{\textstyle{{\scriptstyle #1}\over {\scriptstyle #2}}}}
\def\fft#1#2{{#1 \over #2}}
\newcommand{\be}{\begin{equation}}
\newcommand{\ee}{\end{equation}}
\newcommand{\eq}[1]{(\ref{#1})}
\def\eqs#1#2{(\ref{#1}-\ref{#2})}
\def\det{{\rm det\,}}
\def\tr{{\rm tr}}
\newcommand{\ho}[1]{$\, ^{#1}$}
\newcommand{\hoch}[1]{$\, ^{#1}$}
\def\ra{\rightarrow}
\def\uha{{\hat {\underline{\a}} }}
\def\uhc{{\hat {\underline{\c}} }}

\newcommand{\tamphys}{\it\small Center for Theoretical Physics,
Texas A\&M University, College Station, TX 77843, USA}

\newcommand{\kings}{\it\small Department of Mathematics, King's College,
London, UK}

\newcommand{\sissa}{\it\small International School for Advanced Studies
 (SISSA/ISAS), Via Beirut 2, 34014 Trieste, Italy}

\newcommand{\newton}{\it\small Isaac Newton Institute for Mathematical Sciences,
Cambridge, UK}

\newcommand{\auth}{\large C.S. Chu\hoch{1}, P.S. Howe\hoch{2} and
E. Sezgin\hoch{3\dagger} }

\begin{document}

\hfill{SISSA 2/98/FM}

\hfill{KCL-TH-98-02}

\hfill{CTP TAMU-1/98}

\hfill{hep-th/9801202}

\vspace{20pt}

\begin{center}

{\Large\bf Strings and D-Branes with Boundaries}
\vspace{30pt}

\auth

\vspace{15pt} 

\begin{itemize} 
\item[$^1$] {\small \em
International School for Advanced Studies (SISSA),
Via Beirut 2, 34014 Trieste, Italy}
\item[$^2$] {\small \em
Department of Mathematics,
King's College, London, UK}

\item[$^3$] {\small \em Center for
Theoretical Physics, Texas A\&M University, College Station, TX
77843, USA}
\end{itemize}

\vspace{60pt}

{\bf Abstract}

\end{center}

The covariant field equations of ten-dimensional super D-branes are
obtained by considering fundamental strings whose ends lie in the
superworldsurface of the D-brane. By considering in a similar fashion
D$p$-branes ending on D$(p+2)$-branes we derive equations describing
D-branes with dual potentials, as well as the vector potentials. 

{\vfill\leftline{} \vfill 
\vskip	10pt 
\footnoterule
{\footnotesize \hoch{\dagger} Research supported in part by NSF Grant
PHY-9722090 \vskip -12pt} 
\vskip	10pt 

\pagebreak
\setcounter{page}{1}

%%%%%%%%%%%%%%%%%%%%%%%%%%%%%%%%%%%%%%%%%%%%%%%%%%%%%%%%%%%%%%%%%%%%%%%

\section{Introduction}

%%%%%%%%%%%%%%%%%%%%%%%%%%%%%%%%%%%%%%%%%%%%%%%%%%%%%%%%%%%%%%%%%%%%%%%

In a recent paper \cite{cs}, an open supermembrane ending on an
M-fivebrane was studied. The worldvolume of the M-fivebrane was taken to
be a supersubmanifold, $M$, of the eleven-dimensional target superspace,
$\uM$. The supermembrane action is an integral over a bosonic three
dimensional worldvolume $\S$, with its boundary $\del\S$ embedded in the
supermanifold $M$, such that \be \del \S \subset M \subset {\unM}
\la{ms} \ee

It was shown that the $\kappa$-symmetry of the total action implies (1)
the eleven-dimensional supergravity equations, (2) a constraint on the
embedding of $M$ in ${\unM}$ and (3) a constraint on a modified super
3-form field strength, $H$, on the superfivebrane worldvolume. The
superembedding constraint and the $H$-constraint completely determine the
superfivebrane equations of motion, although the $H$-constraint can be
derived from the embedding constraint in this case. 

In this paper, this result is extended to the following configurations of
branes:

\begin{itemize}
\item[1.] Fundamental type II strings ending on D-branes
\item[2.] Type II D$p$-branes ending on  D$(p+2)$-branes
\end{itemize}

The target space is $(10|32)$ dimensional type IIA or
type IIB superspaces. We use the notation $(D|D')$, where $D$ is the
real bosonic dimension and $D'$ is the real fermionic dimension of a
supermanifold. The embedded supermanifold $M$ has dimension
$(2|16)$ in case (1) and $(p+1|16)$ for case (2). 

In all cases we find that the $\k$-symmetry of the total system implies
the ten-dimensional type IIA or type IIB supergravity equations and a
constraint on the embedding of $M$ in ${\unM}$. In addition, in case (1)
we find a constraint on a modified super 2-form field strength on $M$
defined as
\be
{\cF}=dA- B\ , 
\ee
where $A$ is the super 1-form potential on $M$, and $B$ is the pullback
of the target space NS-NS super 2-form $M$. We will use the same letter
to denote the target space and worldsurface superforms, since it should
be clear from the context whether a pullback is required. The
superembedding constraint and the ${\cF}$-constraint determine
completely the dynamics of the D-brane on which the fundamental string
ends. At the linearised level, these constraints are shown to be
precisely the dimensional reduction of the ten-dimensional Maxwell
superspace constraints.

In case (2) the construction leads naturally to the introduction of a
$p$-form potential on $M$ in addition to the usual one-form potential.
The field strength forms corresponding to these potentials are
esentially dual to one another, so that the dual versions of D-branes
are automatically generated by this method. Again, constraints on the
field strengths are derived and imply the equations of motion for the
D$(p+2)$-brane when the embedding condition is taken into account.

In the case of M-fivebrane, the 3-form $H$ was introduced in
\cite{hs1,hs2} for convenience in describing the field equations and it
was shown that the $H$-constraint is a consequence of the superembedding
condition \cite{hs2,hsw1}. In \cite{hs1}, it was observed that the
analogue of the $H$-constraint arises in the description of various
superbranes and, in particular, the super $D$-branes naturally
accomodate an ${\cF}$-constraint, where ${\cF}$ is a modified two-form
field strength. It was also noted in \cite{hs1} that for certain
superbranes, e.g. the $D8$-brane, the ${\cF}$-constraint is needed to
put the theory on-shell \cite{hrss}.

In the approach presented in \cite{cs}, both the superembedding
condition and the $H$-constraint arise naturally from the requirement of
$\k$-symmetry. Similarly, here we will show that both the superembedding
condition and the ${\cF}$-constraint arise naturally from the
considerations of $\k$-symmetry of suitable open branes ending on
$D$-branes.

%%%%%%%%%%%%%%%%%%%%%%%%%%%%%%%%%%%%%%%%%%%%%%%%%%%%%%%%%%%%%%%%%%%%%%%

\section{Fundamental Type II Strings Ending on $D$-Branes}

%%%%%%%%%%%%%%%%%%%%%%%%%%%%%%%%%%%%%%%%%%%%%%%%%%%%%%%%%%%%%%%%%%%%%%%

In this section, we consider the fundamental type II strings in ten
dimensions with boundary on a D$p$-brane. The string worldsheet is
bosonic. We will take its boundary, however, to lie in a bosonic
submanifold of a supermanifold $M$ of dimension $(p+1|16)$, which in
turn is a submanifold of a target space $\unM$ of dimension $(10|32)$.
We use the notations and conventions of \cite{hs1}. In particular, we
denote by $z^{\unM}=(x^{\unm},\th^{\umu})$ the local coordinates on
$\unM$, and by $\unA=(\una,\ua)$ the target tangent space indices. We
use the ununderlined version of these indices to label the corresponding
quantities on the worldsurface. The embedded submanifold $M$, with local
coordinates $z^M=(x^m,\th^{\m})$, is given as $z^{\unM}(z)$. 

We shall consider type IIA and type IIB superspaces. The fermionic
coordinates consist of two Majorana-Weyl spinors. In type IIA
superspace, these spinors carry opposite chiralities which can be
combined into a single 32 component Majorana spinor, while in type IIB
superspace they are of the same chirality. We will use the fermionic
index $\ua$ in both cases, though in type IIB superspace it is
understood to be a composite index of Majorana-Weyl spinor index and an
$SO(2)$ doublet index, acted on by a direct product of chirally
projected $\C$-matrices and $SO(2)$ matrices. Further
details of our notation and conventions are given in the Appendix.

%%%%%%%%%%%%%%%%%%%%%%%%%%%%%%%%%%%%%%%%%%%%%%%%%%%%%%%%%%%%%%%%%%%%%%%

\subsection{Constraints from $\k$-Symmetry of the Open String}

%%%%%%%%%%%%%%%%%%%%%%%%%%%%%%%%%%%%%%%%%%%%%%%%%%%%%%%%%%%%%%%%%%%%%%%

We consider the following action for the total system of a type II open
string ending on a D-brane (with the target metric taken to be in the
the Einstein frame),

\be
S=-\int_\S d^2 \xi \left (\sqrt{-g} + \e^{ij} B_{ij}\right)
+ \int_{\del\S} d \t A, \la{saction}
\ee

where $\x^i~(i=0,1)$ are the coordinates of the string worldsheet $\S$,
$\tau$ is the coordinate on the boundary $\del\S$. We will take both
ends of the string to lie on a D$p$-brane supermanifold $M$ of
dimension $(p+1|16)$.
%%%
%%%%%%%%%%%%%%%%%%%%%
\footnote{In general, the end points may lie on two different
D$p$-branes or one end-point of a semi-infinite open string may lie on a
D$p$-brane while the other end is feely moving. It is sufficent to
consider the case where both end-points are ending on a single
D$p$-brane for the purpose of deriving the constraints that govern the
dynamics of the D$p$-brane. It is straightforward to generalize the
discussion for the other two cases.}. 
%%%
%%%%%%%%%%%%%%%%%%%%%
$A$ is the pullback to $\del\S$ of a super one-form defined on $M$. The
induced metric $g_{ij}$, and the pullbacks $B_{ij}, A$ are defined as:
\bea
B_{ij} &=& E_j{}^{\unB} E_i{}^{\unA}  B_{\unA\unB}\ ,
\nn\\
A &=& E_{\t}{}^C A_{C}\ ,  
\nn\\
g_{ij} &=& E_i{}^{\una} E_j{}^{\unb} \eta_{\una\unb}\ ,\la{cbg}
\eea
where $\eta_{\una\unb}$ is the Minkowski metric in ten dimensions, and
\bea
E_i{}^{\unA} &=& \del_i z^{\uM} E_{\uM}{}^{\unA}\ ,
\nn\\
E_{\t}{}^A &=& \del_{\t} z^M E_M{}^A \ ,
\la{ee}
\eea
where $E_{\uM}{}^{\unA}$ is the target space supervielbein and $E_M{}^A$
is the worldsurface supervielbein. We note the useful relation
\be
d\xi^i E_i{}^{\unA}|_{\del\S} =  d\t E_{\t}{}^A E_A{}^{\unA}|_{\del\S}\ .
\la{useful}
\ee
where $E_A{}^{\unA}$ is the embedding matrix which plays an important
r\^{o}le in the description of the model. It is defined by
\be
E_A{}^{\unA}=E_A{}^M\del_{M}z^{\unM}E_{\unM}{}^{\unA},
\ee

We consider a $\k$-symmetry transformation of the form 
\bea
\d_{\k} z^{\una} &=& 0 \ , \la{k1}\nn\\
\d_{\k} z^{\ua} &=&  \ft12\,\k^{\uc}(\xi) (1+\C)_{\uc}{}^{\ua} \ ,\la{k2}
\eea
on the string worldsheet $\S$, where 
\be
(\C)_{\ua}{}^{\ub} = \frac1{2\sqrt{-g}}\e^{ij}
\left(\c_{ij}P\right)_{\ua}{}^{\ub}\ , \la{c}
\ee
where 
\be
P=\cases{\C_{11} & (IIA) \cr
       \s_3& (IIB)}
\ee
Note that the $\C_{11}$ acts on a 32 component Majorana index, while
$\s_3$ acts on the $SO(2)$ doublet index of the two 16-component
Majorana-Weyl spinors of same chirality.

The boundary $\k$-transformations will be taken to be of the form
as in \cite{cs}, namely
\bea
\d_{\k} z^{\una} &=& 0\ , \la{bk1}\nn\\
\d_{\k} z^{\ua} &=& \ft12 \k^{\uc}(\s) \left(1+\C_{(p+1)}\right)_{\uc}{}^{\ua} 
\quad\quad \mbox{on $\del\S$} \ , \la{bk2}
\eea
where the matrix $\C_{(p+1)}$ is defined by
\be
E_{\ua}{}^{\a} E_{\a}{}^{\uc} = \ft12 \left(1+\C_{(p+1)}\right)_{\ua}{}^{\uc}\ .
\ee
The matrix $E_{\a}{}^{\uc}$ is obtained from 
$E_{\unA}{}^A$ which is the inverse of $E_A{}^{\unA}$. For more details, see
\cite{hs2,hsw1}. 

The vanishing of the terms on $\S$ imposes constraints on the torsion
super two-form $T$, and the super three-form $H=dB$, such that they are
consistent with the equations of motion of the ten-dimensional type II
supergravities \cite{getal}. 

The constraints which follow from $\k$-symmetry on $\S$ are
\bea
T_{\ua\ub}{}^{\una} &=& -i(\C^{\una})_{\ua\ub}\ , \nn\\
T_{\ua \unb}{}^{\unc} &=& \d_{\unb}{}^{\unc}~\chi_{\ua}\ , \la{tc1}
\eea
and 
\bea
&&
H_{\ua\ub\uc} =0\ ,\nn\\
&&
H_{\ua\ub \una} = i(\C_{\una}Q)_{\ua\ub}\ ,\nn\\
&&
H_{\ua \una\unb} = \left(\C_{\una\unb} Q \chi\right)_{\ua} 
\la{hc1}
\eea 
where 
\be
Q=\cases{\C_{11} & (IIA) \cr
              \s_1& (IIB)}
\ee
and $\chi_{\ua}$ is a spinor superfield proportional to  the
dilaton superfield of the supergravity background. 

The remaining variations are on the boundary. Proceeding exactly as
in \cite{cs}, we learn that they vanish provided that
the following two constraints are satisfied:
\bea
&&E_{\a}{}^{\una}=0, \la{basic}\\
&&{\cF}_{\a B}=0\ .  \la{sf}
\eea
Here,
\be
{\cF}= dA - B\ , \la{ff}
\ee
is the modified 2-form superfield strength which
satisfies the Bianchi identity
\be
d {\cF} = - H\ .  \la{bif1}
\ee
There will also be a mixture of Dirichlet and Neumann boundary
conditions from the requirement that the action be stationary when the
string field equations hold. These can be derived straightforwardly as
in the case of the open supermembrane which has been discussed in detail in
\cite{cs}.

%%%%%%%%%%%%%%%%%%%%%%%%%%%%%%%%%%%%%%%%%%%%%%%%%%%%%%%%%%%%%%%%%%%%%%%

\subsection{Solution of the Linearised Constraints}

%%%%%%%%%%%%%%%%%%%%%%%%%%%%%%%%%%%%%%%%%%%%%%%%%%%%%%%%%%%%%%%%%%%%%%%

In this section, we shall analyse the embedding condition \eq{basic} and
the the ${\cF}$-constraints \eq{sf} in order to extract the equation of
motion for the D-brane worldvolume fields. To determine the field
content, it is sufficient to study the linearised constraints in flat
target space limit. 

The supervielbein for the flat target superspace is, 
\beqa
E^{\una}&=&dx ^{\una} -{i\over2} d\th^{\ua}(\C^{\una})_{\ua\ub}\th^{\ub} \nn\\
E^{\ua}&=& d\th^{\ua}\ .
\eeqa
Let us choose the physical gauge,
\beqa
x^{\una}&=&\cases{x^a &\cr x^{a'}(x,\th)  &\cr} \nonumber\\
\th^{\ua}&=&\cases{\th^{\a}&\cr\th^{\a'}(x,\th)&\cr}
\eeqa
and take the embedding to be infinitesimal so that $E_A{}^M\del_M$ can
be replaced by $D_A=(\del_a,D_{\a})$ where $D_{\a}$ is the flat
superspace covariant derivative on the worldsurface, provided that the
embedding constraint holds. In this limit the embedding matrix is: 
\bea
E_a{}^{\unb}&\rightarrow&\cases{\d_a{}^b &\cr\del_a X^{b'}&\cr}\nn\\
E_{\a}{}^{\unb}
&\rightarrow&\cases{0 &\cr D_{\a}X^{a'}-
       i(\C^{a'})_{\a\b'}\th^{\b'}&\cr}\nn\\
E_a{}^{\ub}&\rightarrow&\cases{0 &\cr \del_a\th^{\b'} &\cr}\nn\\
E_{\a}{}^{\ub}&\rightarrow&\cases{\d_{\a}{}^{\b}&\cr D_{\a}\th^{\b'} &
\cr}\la{lin}
\eea
where
\be
X^{a'}:=x^{a'}+{i\over2}\th^{\a}(\C^{a'})_{\a\b'}\th^{\b'}\ .
\ee

Using this in the embedding condition \eq{basic} we find, at the linearised
level,
\be
D_{\a}X^{a'}=i(\C^{a'})_{\a\b'}\th^{\b'}\ . \la{mc}
\ee

The Bianchi identity \eq{bif1} in component form is, 
\be
D_{[A} {\cF}_{BC]} +T_{[AB}{}^E {\cF}_{|E|C]} = 
- \ft13   E_{C}{}^{\unC} E_B{}^{\unB} E_{A}{}^{\unA} H_{\unA\unB\unC}\ . 
\la{sbi}
\ee
Linearising this equation using \eq{lin}, we find that
the $ABC=(\a\b\c)$ component of this identity is satisfied
automatically,  while the $(ab\c)$, $(abc)$ and
$(\a\b c)$ components give
\bea
D_\c {\cF}_{ab}&=& -2 i \, \del_{[a} \th^{\a'}(\C_{b]} Q)_{\c \a'}
  \la{df1}
\\
\del_{[c}{\cF}_{ab]} &=& 0\ ,\la{df2}
\\
(\c^b)_{\a\b} {\cF}_{bc} &=&  i \, (\C_c Q)_{\a\b}+
			      (\C_{a'}{Q})_{\a\b}\, \del_c X^{a'} +
 		              2 D_{(\a} \th^{\b'} (\C_c{Q})_{\b )\b'}
\la{df3}
\eea
Our strategy is to interpret these equations, together with \eq{mc}, as
the dimensional reduction of the $(N=1)$ ten-dimensional super Maxwell
system to $(p+1)$ dimensions. The relation between ${\cF}$ and the
non-covariant $F=dA$ follows from \eq{ff}. 
In component form, \eq{ff} reads
\be 
{\cF}_{AB} =F_{AB} -  E_B{}^{\unB} E_A{}^{\unA} B_{\unA\unB}.
\ee
These relations are:
\bea
{\cF}_{ab} &=&F_{ab}\ ,\la{ff1}
\\ 
{\cF}_{\a\b} &=& F_{\a\b}\ ,\la{ff2}
\\
{\cF}_{a\a} &=&  F_{a\a}- i (\C_a{Q})_{\a\b'}\, \th^{\b'}\ . \la{ff3}
\eea
Using \eq{sf}, the constraints \eq{df1}, \eq{df2} and  \eq{ff1}-\eq{ff3}
can be summarized as
\bea
F_{\a\b}&=&0\ , \la{ym1}
\\
F_{a\a}&=& i (\C_a{Q})_{\a\b'}~\th^{\b'}\ ,\la{ym2}
\\
D_\c F_{ab} &= & -2 i \, \del_{[a} \th^{\a'}(\C_{b]} Q)_{\c \a'}
\la{ym3}
\\
\del_{[c} F_{ab]} &=& 0\ . \la{ym4}
\eea
It is now easy to combine these and \eq{mc} to obtain ten dimensional
master constraints. To do this, we first define a ten dimensional vector
superfield $A^{\una}$, and a spinor superfield $\l$:\footnote{
$A$ is imaginary due to our choice \eq{tc1} of the torsion. 
}
\bea 
A^a, i X^{a'}  &\rightarrow &A^{\una}\ ,\nn\\
\th^{\a'}  &\rightarrow & \l^\a\ ,
\eea
where the index $\a$ now labels a sixteen component Majorana-Weyl spinor
in ten dimensions. With these definitions, the constraints \eq{mc} and
\eq{ym2} combine to
\be
F_{\una \a}=  (\s_{\una})_{\a\b}~\l^\b\ ,
\ee
where the $\s$ matrices are the ten dimensional chirally projected $\c$-
matrices. This constraint, together with \eq{df3}, \eq{ym1}, \eq{ym3},
and \eq{ym4} are precisely the superspace constraints of ten dimensional
super Maxwell system that satisfy the ten dimensional Bianchi identity
$dF=0$. In particular, the constraint \eq{df3} is the dimensional
reduction of the ten dimensional Bianchi identity
\be
D_{(\a} F_{\b)\unc} + \frac{i}{2} (\s^{\unb})_{\a\b}~F_{\unb\unc}=0\ ,
\ee
for $\unc=c$. The other component of this equation, i.e. for $\unc=c'$,
is also satisfied, thanks to the supersymmetry algebra,
\be
\{ D_{\a}, D_{\b} \} = i \, (\s^{\una})_{\a\b} \del_{\una}.
\ee
In doing these calculations, we
have used the properties of the $\C$-matrices, given in the Appendix, to set
the first term on the right hand side of \eq{df3} equal to zero.

We conclude that the linearised versions of our two master constraints
\eq{sf} and \eq{basic} describe precisely the dimensional reduction of the
ten-dimensional super Maxwell system to D$p$-brane worldvolume. We
expect \cite{hs1,hrss} that the full constraints \eq{basic} and \eq{sf} imply
the full field equations that follow from the super D-brane actions of
\cite{d0,d1,d2,d3,d4}.

%%%%%%%%%%%%%%%%%%%%%%%%%%%%%%%%%%%%%%%%%%%%%%%%%%%%%%%%%%%%%%%%%%%%%%%

\section{D$p$-brane ending on a D$(p+2)$-brane}

%%%%%%%%%%%%%%%%%%%%%%%%%%%%%%%%%%%%%%%%%%%%%%%%%%%%%%%%%%%%%%%%%%%%%%%

In this section we study an open D$p$-brane ending on a D$(p+2)$-brane
and show that it naturally gives rise to a dual potential on the
worldsurface of the $(p+2)$-brane $M$.

%%%%%%%%%%%%%%%%%%%%%%%%%%%%%%%%%%%%%%%%%%%%%%%%%%%%%%%%%%%%%%%%%%%%%%%

\subsection{The Action and $\k$-symmetry Constraints}

%%%%%%%%%%%%%%%%%%%%%%%%%%%%%%%%%%%%%%%%%%%%%%%%%%%%%%%%%%%%%%%%%%%%%%%

The action for a type II $\k$ symmetric D-brane has been constructed in
\cite{d0,d1,d2,d3,d4}. $\k$ symmetry implies constraints for the target
superspace torsion, the NS-NS three form field strength and the RR
field strength. For an open D$p$-brane ending on a D$(p+2)$-brane, we
propose the action (here we take the target metric to bein the string
frame)
\be
S= \int_{\S} \left( -e^{-\phi}\sqrt{-\det (g_{ij} + {\cF}_{ij})} +
Ce^{\cF} +m \o_{p+1} \right) +\int_{\del \S} A_p\ , \la{action}
\ee
where
\be
{\cF} =dA - B \ , \la{fab}
\ee
and $\o_{p+1}(A,dA)$ is the Chern-Simons form present for even $p$ in a
massive IIA background, with $m$ being the mass parameter. We define
this form by the relation 
\be
d\o_{p+1}(A,dA)= (e^{dA})_{p+2}\ .
\ee
In \eq{fab}, $B$ represents the pullback of the target space super
two-form to the {\it bosonic} D$p$-brane worldvolume. The potential
$A_p$ is identified with the pullback onto the bosonic boundary
$\partial\S$ of a $p$-form potential living on the D$(p+2)$-brane
superworldvolume. Furthermore, it is assumed that the pullback of the
field strength ${\cF}_2=dA_1-B$ defined on the superworldvolume of the
D$(p+2)$-brane onto $\partial\S$ coincides with ${\cF}$ for the
$Dp$-brane restricted to the (bosonic) boundary. Thus we have both a
one-form potential $A_1$ and the dual $p$-form potential $A_p$ on the
$(p+2)$-brane. For simplicity, we will use in the following the same
symbol $\cF$ to denote both the two-form field strengths on the
D$p$-brane and on the D$(p+2)$-brane. It should be clear from the
context which one is being referred to.

For later reference, we record here the definitions of the target space
RR field strengths:
\be
G= dC-CH + m e^B\ ,
\ee
where 
\be
H=dB\ .
\ee
We also record the Bianchi identity
\be
dG=GH \ , \la{big}
\ee
Note that the $m$-dependent terms have cancelled. We use the superspace
conventions of \cite{howe} according to which the exterior derivative
acts from the right. 

The field strenghts ${\cF},G,H$ are invariant under the gauge
transformations
\be
\d A=\l\ , \quad\quad \d B=d\l\ ,\quad\quad 
\d C=-m \l\,e^B \ , \la{g1}
\ee
where $\l$ is a target space super one-form gauge parameter. The field
strength $G$ is also invariant under the gauge transformation
\be
\d A=0\ , \quad\quad \d B=0\ ,\quad\quad \d C= e^B\,d\m\ . \la{g2}
\ee
where $\m$ is a target space superform of appropriate rank. 

Since the gauge variation of the Chern-Simons form has the form
\be
\d \o_{p+1}= \l\,e^{dA} + dX^1_p\ ,
\ee
for some $p$-form $X^1_p(\l,A,dA)$ defined by this equation, the action
\eq{action} is invariant under the gauge transformations \eq{g1} and
\eq{g2}, provided that $A_p$
transforms as
\be
\d A_p= -m\,X^1_p -\m\,e^{dA}\ .
\ee

Next, we turn to the discussion of $\k$-symmetry. One can verify that,
under a $\k$-symmetry transformation, the vanishing of the variations on
$\S$ impose constraints on the supertorsion $T$, the NS-NS field
strength $H$ and the RR field strengths $G$ \cite{d0,d1,d2,d3,d4} such
that they are consistent with the field equations of the type II
supergravities.

The remaining variations are on the boundary, and they take the form 
\be 
\delta S = \int_{\del \S} i_{\k} {\cF}_{p+1},
\la{dS}
\ee
where the modified field strengths ${\cF}_{p+1}$ for the worldvolume
potentials $A_p$ are 
\be 
{\cF}_{p+1} := d A_p + (C e^{\cF})_{p+1} 
+ m\o_{p+1}\ , \la{Ft}
\ee
They satisfy the Bianchi identity
\be
d {\cF}_{p+1} = (G\,e^{\cF})_{p+2}\ .  
\la{bif}
\ee
Observe that the $m$-dependent terms have cancelled. Using $i_{\k} {\cF}=0$,
the vanishing of \eq{dS} implies that 
\be
{\cF}_{\a B_1 \cdots B_p}=0\ . \la{mfc}
\la{cft}
\ee
In addition, we must have the usual embedding constraint
\be 
E_{\a}{}^{\una}=0\ , \la{mc2}
\ee 
by similar argument to the one given in the preceeding discussion of
string ending on branes.  

%%%%%%%%%%%%%%%%%%%%%%%%%%%%%%%%%%%%%%%%%%%%%%%%%%%%%%%%%%%%%%%%%%%%%%%

\subsection{An Example: The D2-brane ending on a D4-brane}

%%%%%%%%%%%%%%%%%%%%%%%%%%%%%%%%%%%%%%%%%%%%%%%%%%%%%%%%%%%%%%%%%%%%%%%

To illustrate the general formalism introduced above we consider an open
D2-brane ending on a D4-brane. According to the results of the
preceeding section there will be two potentials, the usual one-form
potential $A_1$, and the dual two-form potential $A_2$, on the D4-brane.
The Bianchi identities are (although the mass papameter $m$ does not
arise in the Bianchi identities \eq{big} and \eq{bif}, we will
nonetheless set it to zero for simplicity)
\be
d{\cF} = -H\ , \la{dfh}
\ee
and
\be
d{\cF}_3=G_4 + G_2 {\cF}\ . \la{dfg}
\ee
In addition we are required to take all the components of both ${\cF}$
and ${\cF}_3$ to be zero, except for those which have solely vectorial
indices. 

In the case of the D4-brane the embedding constraint is enough to force
the equations of motion \cite{hs1,hrss} which are usually written in
terms of the one-form potential $A_1$. However, there is also a dual GS
version \la{pkt} in which one replaces $A_1$ with a two-form $A_2$.
Since in the superembedding formalism the brane is on-shell due to the
basic constraint \eq{mc2} it follows that we should be able to construct
either, or indeed both, versions, and the open brane set-up naturally
gives both potentials.

To analyse the above Bianchi identities we set
\be
E_{\a}{}^{\ua}=u_{\a}{}^{\ua} + h_{\a}{}^{\b'} u_{\b'}{}^{\ua}\ ,
\ee
and
\be
E_{a}{}^{\una}=u_a{}^{\una}\ ,
\ee
where $u$ denotes an element of $Spin(1,9)$, in either the spin
representation or the vector representation according to the indices.
The basic constraint \eq{mc2} implies that 
\be 
E_{\a}{}^{\ua}
E_{\b}{}^{\ub}T_{\ua\ub}{}^{\unc}= T_{\a\b}{}^{c} E_c{}^{\unc}\ ,
\ee 
from which one finds that \cite{hrss}
\bea 
h_{\a}{}^{\b'}\rightarrow h_{\a i}{}^{\b'j}
&=& i\d_i{}^j(\c^{ab})_{\a}{}^{\b'}h_{ab} \ ,\nn\\
&&\nn\\
T_{\a i\b j}{}^{c} &=& -i\h_{ij}\left((\c^b)_{\a\b}m_b{}^c + 
C_{\a\b}m^c \right)\ . 
\eea 
Here we have introduced the two-step notation for the spinor indices on
$M$. The index $\a$, running from 1 to 16, is rewritten as the pair $\a
i$, where $\a$ is a five-dimensional spinor index (running from 1 to 4)
and $i$ is an $Sp(4)$ index, also running from 1 to 4. The $m$-tensors
are given by \cite{hrss}
\bea 
m_{ab} &=& (1-2y_1)\h_{ab} + 8(h^2)_{ab}\ , \nn\\
&&\nn\\
m^a &=& -i\e^{abcde} h_{bc} h_{de} \ ,
\eea 
where $y_1$ and $y_2$ denote the two invariants 
\bea 
y_1 &=& {\rm tr}\ h^2\ , \nn\\
y_2 &=& {\rm tr}\ h^4\ . 
\eea

It is straightforward to check the Bianchi identities for the ${\cF}$'s
using this information. If we take the target space to be flat for
simplicity, the only non-vanishing components of the RR tensors $G_4$
and $G_2$ are 
\be 
G_{\ua\ub\unc\und}= -i(\C_{\unc\und})_{\ua\ub} \ ,
\ee 
and
\be 
G_{\ua\ub}=-i(\C_{11})_{\ua\ub}\ , 
\ee 
while the non-vanishing component of the NS tensor $H$ is 
\be
H_{\ua\ub\unc}=-i(\C_{\unc}\C_{11})_{\ua\ub}\ . 
\ee 
The dimension zero component of the ${\cF}$ Bianchi \eq{dfh} is found to be
satisfied if \cite{hrss} 
\be
m_a{}^c {\cF}_{cb}= 4h_{ab} \ ,
\ee 
which can be rewritten as 
\be 
{\cF}_{ab}={4\over
(1+4y_1^2-16y_2)}((1+2y_1)h_{ab}-8(h^3)_{ab})\ . 
\ee 
In obtaining this result, it is useful to note the identity $ X^5=
\ft12 X^3\tr X^2-\ft18 X(\tr X^2)^2 +\ft14 X\tr X^4$, which holds for
any $5\times 5$ matrix $X$. One also finds that the dimension zero
component of the $F_3$ Bianchi identity \eq{dfg} is satisfied if 
\be 
\tilde
{\cF}_{ab}={4\over (1+4y_1^2-16y_2)}((1-2y_1)h_{ab} + 8(h^3)_{ab})\ ,
\ee 
where $\tilde {\cF}_{ab}$ is the dual of ${\cF}_{abc}$, 
\be 
\tilde
{\cF}_{ab}=-{1\over 3!}\e_{abcde} {\cF}^{cde}\ .
\ee 
It is in fact enough to show that the dimension zero components of these
identities are satisfied to show that that the complete identities are.
Furthermore, we know from \cite{hs1,hrss} and from the string discussion
that the worldsurface multiplet for a D4-brane with a one-form potential
satisfying the standard constraints is on-shell. It is easy to confirm
that this is still the case here by considering the linearised limit in
which it becomes clear that ${\cF}_{abc}$ is the dual of ${\cF}_{ab}$.
Since the three-form field strength is not a new independent field the
version we have derived here is also on-shell. 

%%%%%%%%%%%%%%%%%%%%%%%%%%%%%%%%%%%%%%%%%%%%%%%%%%%%%%%%%%%%%%%%%%%%%%%

\section{Comments}

%%%%%%%%%%%%%%%%%%%%%%%%%%%%%%%%%%%%%%%%%%%%%%%%%%%%%%%%%%%%%%%%%%%%%%%

In this paper we have shown that the equations describing various branes
in superspace can be derived by considering the $\k$-symmetry of an open
brane of appropriate dimension ending on the brane of interest. It is
remarkable that the $\k$-symmetry considerations for open superbarens,
within the framework of superembeding approach, give rise to dual
formulations of $D$-branes automatically. Traditional methods to derive
the dual $D$-brane actions have been considered in \cite{sch}. Here, we
find that the potential dual to the usual Born-Infeld vector is
furnished in a natural fashion by a suitable $p$-form that lives on the
boundary of an open $D$-brane which supports its own Maxwell field.

The results presented here, in our opinion, also furnish further
evidence for the power of superembedding approach to a geometrical and
elegant description of all superbranes. Indeed, this approach should be
applicable to the description of type IIA/B solitonic fivebranes and
type I strings/fivebranes as well. It would also be interesting to
consider a limit of the model considered here to extract an action for
self-dual string in six dimensions. Yet another possible application
would be a description of longer superembedding chains or brane
networks. Results in this direction will be reported elsewhere
\cite{chs}. 

The formalism described here is a hybrid one involving bosonic
worldsurface of the first brane but the superworldsurface of the brane
to be investigated. One may envisage an approach in which the open brane
worldsurface is also elevated into a superspace. This would make the
target space and worldsurface supersymmetry manifest and moreover in
this approach the geometrical meaning of $\k$-symmetry as odd
diffeomorphisms of the superworldsuface would become more transparent.
Indeed a purely superspace description of open superbranes ending on
other superbranes is possible, as we will be shown elsewhere \cite{chs}.

\bigskip\bigskip

\noindent{\large \bf Acknowledments}

\bigskip

We thank Per Sundell for helpful discussions. One of the authors (E.S.)
thanks the Abdus Salam International Center for Theoretical Physics in
Trieste for hospitality.

\bigskip\bigskip

%%%%%%%%%%%%%%%%%%%%%%%%%%%%%%%%%%%%%%%%%%%%%%%%%%%%%%%%%%%%%%%%%%%%%%%

\noindent{\Large\bf Appendix}

%%%%%%%%%%%%%%%%%%%%%%%%%%%%%%%%%%%%%%%%%%%%%%%%%%%%%%%%%%%%%%%%%%%%%%%

\bigskip

Here we collect the properties of the various $\C$ matrices in diverse
dimensions. 

For type IIA, we use the conjugation matrices,

\be
C_{\ua\ub}  = \cases{-i\s_2\times C\times \eta\ , \ \ \ \ p=0,4,8 \nn\cr
                         1    \times C\times \eta\ , 
\ \ \ \ \ \ \ \ \  p=2,6   \cr}
\ee
and the $\C$-matrices 
$$
(\C^{\una})_{\ua}{}^{\ub} = \cases{
\s_1\times  \c^a \times 1    \cr
\s_3 \times 1 \times \c^{a'} \ \ \ \ \ \ p=0,4,8 \cr} 
$$
\medskip
\be
(\C^{\una})_{\ua}{}^{\ub} =\cases{
\s_3\times  \c^a \times 1    \cr
\s_1 \times 1 \times \c^{a'} \ \ \ \ p=2,6\cr} 
\ee

where $\c^a$ and $\c^{a'}$ are the $\c$-matrices, while $C$ and $\eta$
are the charge conjugation matrices of $SO(p,1)$ and $SO(9-p)$,
respectively. The matrices $\c^a$ and $\eta$ are symmetric for $p=0,2,8$
and antisymmetric for $p=4,6$, while the matrices $C$ and $\c^{a'}$ are
symmetric for $p=0,6,8$ and antisymmetric for $p=2,4$.

The chirality matrix $\C_{11}=\C_0\C_1\cdots \C_9$ is given by
\be
(\C_{11})_{\ua}{}^{\ub} =  (\s_2 \times 1 \times 1)_{\ua}{}^{\ub}\ .
\ee

A 32 component Majorana spinor $\psi$  in ten dimensions decomposes as
\be
\psi^{\ua} =\left( \ba{c} \psi^\a \\ \psi^{\a'} \ea \right) \la{psi}
\ee
where $\a=1,...,16$ labes the fermionic coordinates of the worldvolume,
and $\a'=1,...,16$ labels the fermionic transverse directions. The $\s$-
matrix factors of the $\C$-matrices act on the doublet \eq{psi}. Thus,
we have for example, 
\be
\C^a_{\a\b'}=0\ , \quad\quad \C^a_{\a'\b'}=-C_{\a\b}\eta_{ij}\ ,  
\quad\quad \C^{a'}_{\a\b'}= C_{\a\b}\c^{a'}_{ij} \nn\\
\ee
where $i=1,...,9-p$ label the vector representation of the transverse
$SO(9-p)$. 

In the case of type IIB $\C$-matrices, we suppress the ten dimensional
$SO(2)$ doublet index, and consider the chirally projected $16\times 16$
$\C$-matrices. The unprimed spinor index labelling the fermionic
worldvolume coordinates, and the primed spinor indices labelling the
transverse fermionic directions are defined by using the projection
operators
\be
P_\pm = \ft12 (1 \pm \s_3) \ ,
\ee
acting on the $SO(2)$ indices $I,J=1,2$, as follows:
\bea
\psi_\a  &=& (P_+\psi)^\a\nn\\
&&\nn\\
\psi_{\a'} &=& (P_-\psi)^{\a'} \ .
\eea
Now we can construct the ten dimensional $\C$-matrices as:
\bea
p=9:  \quad\quad  \C^{\una}_{\ua\ub} &=& \c^a_{\a\b}~P_+ \nn\\
&&\nn\\
p=7:\quad\quad \C^{\una}_{\ua\ub} &=& \cases{ \s^a_{\a\db}~P_+ \cr
   C_{\a\b}~P_-\cr}  \nn\\
p=5:\quad\quad \C^{\una}_{\ua\ub} &=& \cases{ \c^a_{\a\b}~\eta_{ij}~P_+ \cr
  \d_\a^\b~ \c^{a'}_{ij}~P_- \cr} \nn\\
&&\nn\\
p=3:\quad\quad \C^{\una}_{\ua\ub} &=& \cases{\s^a_{\a\db}~ \d_i^j~P_+ \cr
  C_{\a\b}~ \c^{a'}_{ij}~P_-\cr} \nn\\
&&\nn\\
p=1:\quad\quad \C^{\una}_{\ua\ub} &=& \cases{\c^a_{\a\b}~\d_i^j~P_+ \cr
     C_{\a\b}~\c^{a'}_{ij}~P_- \cr} 
\eea
where $\c_{\a\b}$ and $\s_{\a\db}$ are the chirally projected
$\c$-matrices appropriate to $(p+1)$-dimensions. For $p=9$, there is no
transverse direction, and consequently $\una=a=0,1,...,9$. In the case of
$p=7$, the two transverse coordinates have been combined into a single
complex coordinate. Further details can be found in \cite{hrss}.

\pagebreak

%%%%%%%%%%%%%%%%%%%%%%%%%%%%%%%%%%%%%%%%%%%%%%%%%%%%%%%%%%%%%%%%%%%%%%%

\ed